\documentclass[conference]{IEEEtran}

\usepackage{graphicx}
\usepackage{color}
\usepackage{soul}
\usepackage{todonotes}
\usepackage{hyperref}
\usepackage{amsmath}
\ifCLASSINFOpdf
\else
\fi

\begin{document}

\title{A Secure Data Enclave and Analytics Platform for Social Scientists}

\newcommand{\NAMENS} {\textsc{Cloud Kotta}} 
\newcommand{\NAME} {\textsc{Cloud Kotta }}

\author{\IEEEauthorblockN{Yadu N. Babuji, Kyle Chard, Aaron Gerow, Eamon Duede}
\IEEEauthorblockA{Computation Institute\\
University of Chicago and Argonne National Laboratory\\
\texttt{\{yadunand,chard,gerow,eduede\}@uchicago.edu}}
}


%


\maketitle

\begin{abstract}
Data-driven research is increasingly ubiquitous and data itself is 
a defining asset for researchers, particularly
in the computational social sciences and humanities. Entire
careers and research communities are built around 
valuable, proprietary or sensitive datasets. 
However, many existing computation resources fail
to support secure and cost-effective storage of data while also
enabling secure and flexible analysis of the data. 
To address these needs we present \NAMENS, a cloud-based architecture
for the secure management and analysis of social science data.
\NAME leverages reliable, secure, and scalable cloud resources
to deliver capabilities to users, and removes the need
for users to manage complicated infrastructure.
\NAME implements automated, cost-aware models for 
efficiently provisioning tiered storage and automatically scaled
compute resources.
\NAME has been used in production for several months and currently
manages approximately 10TB of data and has been used to process 
more than 5TB of data with over 75,000 CPU hours.  
It has been used for a broad variety of text analysis 
workflows, matrix factorization, and various machine learning
algorithms, and more broadly, it supports fast, secure and cost-effective
research.

\end{abstract}


%
\IEEEpeerreviewmaketitle

\section{Introduction}



Data is fast becoming a crucial, defining, asset for researchers.
Entire fields, including those new to computational
practices, are quickly embracing data-driven research. 
However, the increasing scale and complexity of analysis and the fact that 
datasets are often proprietary, or sensitive, creates unique new challenges.
The centrality of data has inspired new processes that are designed for specific datasets,
which typically results in tightly coupled environments that discourage reusability and agility. 
To support the needs of data-driven research, we developed \NAMENS\footnote{Available at \url{https://github.com/yadudoc/cloud\_kotta}.}, a unique cloud-based
framework that enables the secure and cost-effective
management and analysis of large, potentially sensitive datasets.

To address the growing reliance 
on data in research (particularly in the social sciences and humanities) scholars are increasingly replacing
on-premise infrastructure with cloud-based solutions such as those 
offered by 
Amazon Web Services (AWS). 
This trend is not difficult to explain: cloud platforms provide 
high reliability, availability, and download performance 
without encumbering researchers with managing
on-site infrastructure. The adoption of cloud-based services 
has also afforded new avenues for exploration.
For example, when storage is co-located with elastic 
computing capacity with which data can be analyzed, 
aggregated, and integrated on-demand, researchers 
can take bigger risks and explore new analyses more flexibly. 
\NAME enables this kind of agility across fluid groups
while also ensuring scalability, security, and data provenance. 

Cloud-based infrastructure also has the advantage of 
helping centralize disparate teams. For example, 
given the sensitivity, value, and size of many datasets, 
it is often not feasible to replicate and download entire datasets
for analysis on local computers or clusters. In many cases, 
circuitous approval procedures are necessary to gain access to 
data, and users must adhere to strict data-use agreements. 
Migrating data from the environment where it is hosted adds 
further complexity in this respect, and cloud-based strategies can 
exacerbate these challenges. 
So, when it comes to cloud-based infrastructure, it is important for researchers to develop a unified strategy. 
\NAME offers a strategy that is cost-effective, secure with 
respect to data policies, offers sustainable short- to long-term storage,
and is scalable for demanding workloads. 

\NAME is designed to address the requirements
of two canonical use cases: managing community datasets securely and
providing scalable compute resources. The first use case is motivated by a growing need for 
researchers to make valuable datasets
available to certain research communities. This might be required
by funding agencies or institutions, though, it is generally helpful 
to share data when establishing new research groups. 
The most important requirements for this use case are that \NAME be:

\begin{itemize}
	\item Secure: Data should be stored securely and accessible only to
	authorized and authenticated users. 
	\item Scalable: Data can be large, the storage system should scale
	to the data.
	\item Reliable: Data should be stored reliably, using 
	backups in case of failure or corruption.
	\item Available: Data should be available to 
	geographically distributed users.
	\item High performance: Large amounts of data should be moveable easily and quickly for analysis, download, and archival.  
	\item Cost-effective: Costs associated with data storage should
	be minimized to encourage use.
\end{itemize}

The second use case is motivated by a large-scale movement towards
data-intensive research. As data sizes grow and analyses
become more computationally intensive, the requirements often exceed the
computational capabilities of individual researchers. As such, 
researchers need to be able to scale analyses from individual
computers to distributed, parallel modes of computing. To address
these priorities, \NAME should be:

\begin{itemize}
	\item Secure: Authorizations should control what data can be analyzed and which analyses should be isolated. 
	\item Scalable: Analyses should scale to data, exploit
	parallelism where possible, and leverage large scale computing infrastructure
	for efficient performance. 
	\item Cost-effective: Costs should be comparable or lower than
	using local compute resources.
	\item Easy to use: Interfaces should make it simple to access the 
	underlying infrastructure.
\end{itemize}

\section{Architecture \& Implementation}
\label{sec:architecture}

\begin{figure}
  \center
  \includegraphics[width=0.45\textwidth]{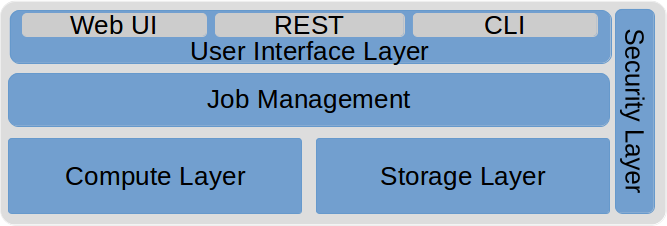}
  \caption{Logical Architecture of \NAME}
    \label{fig:logical_arch}
\end{figure}

The \NAME architecture is depicted in \figurename~\ref{fig:logical_arch}.
The entire system is comprised of a web interface and REST API for accessibility;
a set of event-based management and monitoring software to ensure reliable job execution;
a compute layer that provides cost efficient compute resources;
a fast and cost-effective storage layer;
and an extensible and customizable security fabric that permeates all of the above components.

\NAME is designed to be deployed on \textit{Amazon Web Services} (AWS), the ecosystem of its intended users.
Where possible, \NAME leverages existing cloud services as they are scalable, reliable, secure, and cost-effective.
The entire \NAME system is open source and can be deployed using a reproducible \textit{CloudFormation} configuration which can be further
customized to match target workloads.

\subsection{User Interface}
\label{sec_interface}
\NAME offers three interfaces:
a web interface, a REST API, and a command line interface (CLI) accessible from the login node.
This range of interfaces supports broad usage scenarios, enabling intuitive web access
for web-based users and advanced programmatic and CLI support to facilitate customizable
and automated invocation by technical users.

The supported interfaces support the same operations including, for example, browse datasets, upload new data, view and download
results from previous analyses, submit and manage analyses.
As with the entire \NAME architecture, the interfaces are secured, restricting access to
authenticated, authorized users.
The general architecture is centered around data stored in AWS Simple Storage Service (S3) buckets. 
Users can browse accessible data that they are permitted to access in S3, and they can also
upload files to their own private S3 buckets.
Once uploaded, files are available to be specified as inputs to submitted jobs.
To support dynamic sharing scenarios, such as
emailing colleagues the results of an analysis, \NAME provides support
to construct short-term, anonymous URLs.

Submitting an analysis requires a description of the application (scripts, executables, etc), a list of inputs (S3, external URLs),
a list of output files to be saved, and a maximum wall-time. 
In addition to supporting jobs with arbitrary executables, applications can be templated to create pipelines with simplified user interfaces.
That is, other users wishing to re-use an analysis can simply complete a customized form with the required parameters.
Users submit jobs via web forms in the Web interface, or by specifying the job as a JSON file for the CLI and REST interfaces.


\subsection{Storage Layer}

\NAMENS's storage layer uses a mix of AWS storage services that provide different guarantees
regarding access time, durability, and availability with different cost models. 
The types of storage used by \NAME are:

\begin{itemize}
  \item \textbf{Elastic Block Storage (EBS)}: a high performance block storage model that can be mounted as a file system on an EC2 instance.
  \item \textbf{S3 \textit{standard}}: a reliable object store that provides high performance access via HTTP(S).
  \item \textbf{S3 \textit{infrequent access}}: an object store with reduced storage cost at the expense of availability.
  \item \textbf{Glacier}: an archival storage model that provides high durability at a low price with high data retrieval times.
\end{itemize}

\begin{figure}
  \center
  \includegraphics[width=0.45\textwidth]{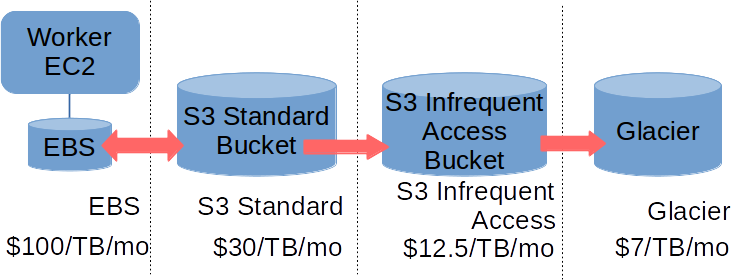}
  \caption{Storage tiers in \NAME and the heuristics used to minimize storage costs. Some costs may differ across regions and configurations.}
    \label{fig:storage}
\end{figure}

\NAME utilizes a caching model that is implemented using automated data life-cycle policies that manage data migration between storage tiers based on access patterns.
Frequently accessed data resides on S3, as it is fast and highly available, whereas data that is
accessed infrequently is moved to Glacier, as it provides durable, low cost storage at the expense of longer retrieval time.
\figurename~\ref{fig:storage} illustrates the storage tiers used in \NAMENS's data model.
Transferring data to lower tiers helps minimize the cost associated with providing high availability.
The primary store for data in \NAME is S3. When data is analyzed, it can either be staged directly from S3 to ephemeral instance storage or EBS (which is subsequently mounted by an instance).
Similarly, archived data stored in Glacier or S3 infrequent access buckets is staged to S3 when needed before being staged for analysis.
Outputs are staged back to S3, guaranteeing durability.

\NAMENS's data model has important advantages over a static storage configuration.
While EBS provides low-latency access, it is more than three times as expensive as S3-Standard and, in addition,
must be mounted as a file system on a live machine to access data.
By storing data in S3, a small overhead is incurred to stage data for compute, however, this latency is nominal in most cases
and comprises a fraction of the time it takes to provision and execute a job.
In addition to the cost benefits, S3 provides support for rich access control features, as well as the ability to publish and access data directly via HTTP.

\subsection{Compute Layer}
The analyses for which \NAME is designed often comprise independent, long running, loosely coupled jobs.
To support this class of workload, \NAME offers a scalable compute layer built upon elastic pools of Elastic Compute Cloud (EC2) instances.

EC2 is a virtualized computing environment in which users can lease virtual machines (VM) with varying computational resources.
Instances are organized by region and \textit{Availability Zone} (AZ).
Regions represent different geographic locations whereas AZs are datacenters located within a specific region.
Pricing models differ between AZs and, due to the independent failure models, users can achieve high 
reliability by distributing applications across AZs. 
EC2 instances are provisioned according to a market model in which users pay for the resources consumed.

\NAME can be configured to use two different EC2 market models. \textbf{On-demand} instances are offered at a fixed hourly price
where instances live until they are terminated by a user. \textbf{Spot} instances are offered using a dynamic price model
where users ``bid'' a maximum hourly price and instances are terminated by AWS if the market price exceeds the user's bid.
Spot markets are typically a fraction of the on-demand price. 

\NAME is designed to support two classes of workloads: short development jobs requiring quick responses,
with minimal compute resources, and longer running production tasks that are computationally intensive, but more tolerant of delays.
To meet the needs of these two asymetric workloads, \NAME offers two independent pools of compute resources. 
The development pool is comprised of on-demand instances, with at least one instance accessible at all times. 
To minimize the cost of executing computationally intensive, long running, yet delay tolerant production workloads, 
we utilize spot instances. \NAME relies on an automated bidding model to provision resources across AZs (to avoid price fluctuations in an AZ). 
Administrators can configure the bidding model to use static or policy-based bid prices (some fraction of the equivalent on-demand price, for example).

While using spot instances can significantly reduce costs, instance revocations are inevitable.
To mitigate the problems associated with instance termination, \NAME manages queues
that ensure jobs that do not complete are resubmitted to the queue and executed again on a new instance. 
By provisioning spot instances on-demand, \NAME can meet the demands of what are often sporadic and bursty
workloads while also helping minimize costs~\cite{chard2015cost}.
\NAME currently uses a pre-defined EC2 instance type for each of it's queues.
In future work, we will integrate \NAME with cost-aware provisioning~\cite{chard2015costworkloads, chard2015cost}
and profiling~\cite{chard2016profiling} approaches to improve the selection of instance types based on cost and execution time. 


\subsection{Job Management}
\NAME uses a job management layer to control the execution of arbitrary user analyses on the compute layer. 
User submitted tasks include the input files, execution scripts, and output files. Users must also 
choose if the task is a development or production job.  
When tasks are submitted, the description is stored in a database such that it can be accessed by the 
job management layer and the instances executing the analysis. 
To execute the task, the job management layer determines the user's access permissions, associates
the user's role with the description, and places the job in the appropriate queue. 

We leverage a queue model as it provides a reliable method for distributing jobs across a pool of EC2 instances.
Worker nodes (EC2 instances) poll the queue for waiting tasks.
If a task is available the worker moves it from a pending queue to the active queue. This active queue is used 
to manage execution and ensure that no tasks are lost. 
The worker retrieves the task description from the database and begins execution. 
In the case where spot instances are used, we must account for unreliability of the underling infrastructure. 
To do so, the job management layer includes a monitoring service that 
periodically checks instance health (e.g., for termination or other failures). If instances are terminated, the monitoring
service will resubmit the task to the pending queue.
Throughout execution, the worker node writes job status markers to the database.
This information provides worker statistics (CPU, I/O and RAM utilization) and job progress, both of which are accessible to the user
to monitor job execution. 
Upon completion, the worker will stage output data to S3, and update the database with the completion code of the task.

\subsection{Security}

The final layer of \NAME is the security fabric that permeates the system. 
\NAME uses Amazon's OAuth~2 model (\emph{Login with Amazon}) for authentication. 
Users are able to login using their Amazon credentials and \NAME is therefore not 
responsible for managing user passwords. 
Before being granted access to the system, users must first be registered in \NAMENS's 
database and given an appropriate role.  
When users login using the OAuth~2 workflow, \NAME is given a short-term delegated access token.
This token can be used to retrieve information about the user from Amazon as well as to use
services as the user. 

\NAME is built around a role-based access control model in which users are assigned roles, 
for example \emph{kotta-public-only} and \emph{kotta-read-WOS-private}, where \emph{WOS} refers to the private Web of Science dataset.
Policies associated with roles define permissions for specific resources (e.g., data access in S3).
Given that all access is controlled by roles, worker nodes must assume a role before they can access restricted data.
Other \NAME services are also given appropriate privileges by internal roles such as \emph{web-server}, \emph{task-executor}.
These roles, unlike user roles, have access to the internal database, queues, notification systems and are capable of controlling scaling functionality.
This role based access model limits validity of credentials to a small window limiting the risk of exposing valuable long-lived credentials.
To adhere with the principle of least privilege, \NAME users are initially given no roles or privileges. They are incrementally granted permissions when required.

Data authorization and access control is implemented on S3 buckets. 
By default, access is not permitted unless it is explicitly granted via a policy.
S3 buckets are associated with policies that prescribe  permissions. 
Policies are then associated with roles that give  permissions to users.
The data stored on S3 buckets are server-side encrypted and accessible only from a Virtual Private Cloud (VPC) Endpoint.
This guarantees that traffic between the S3 bucket and the compute instances remain private.
Output data is also stored in S3. It is initially created as a private object only accessible to the creator. 
As \NAME is used by collaborating groups of users, it is important that data can be shared. 
To provide this capability, we use short-term signed URLs, like those used by other Cloud services (e.g., Google Drive and DropBox)
that provide short term access to the holder of the URL. 

\NAME implements a strong security model between instances and other services. The compute
layer is hosted within a private subnet enclosed within a VPC. This ensures that compute
instances are not directly accessible via the internet.  
Worker nodes are associated with a \emph{task-executor} role that has few privileges (e.g., it cannot access any data)
However, this is a trusted role that can be used to change to a user role for a short period of time. 
This is crucial as it enables a worker node, executing on behalf of a user, the ability to inherit the user's role
and therefore access any data needed by the job. This approach ensures that even 
a running task is only able to access data for which the user is authorized to access.
After staging data, the worker returns to the \emph{task-executor} role to execute the job. 

Finally, \NAME records every action performed by the system to ensure
that data access and usage can be thoroughly audited. 
This information is recorded in a database such that administrators
can export an audit log for any dataset, user, or service.





\section{Usage \& Applications}
\label{sec:applications}

\NAME has been deployed and used over the past six months by a range of 
computational social scientists. 
Active use cases for \NAME
include text analysis, semantic word embedding, matrix factorization, 
optical character recognition, and social network analysis. 
Here we describe usage and 
illustrate some representative analyses.

\subsection{Usage}
\label{sec:usage}

\begin{figure}
	\centering
    \includegraphics[width=0.475\textwidth] {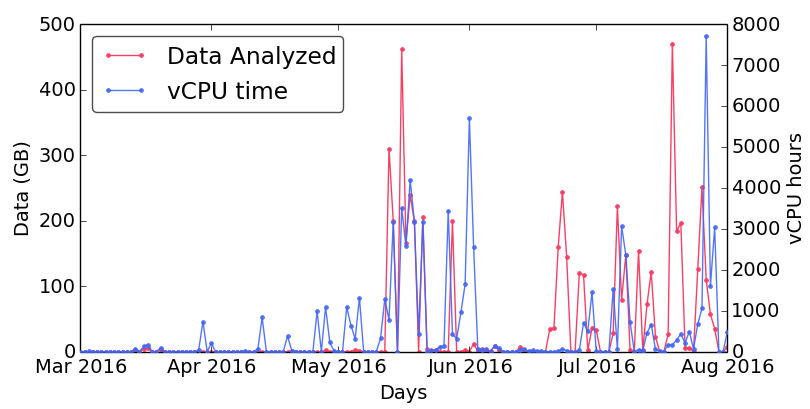}
	\caption{Monthly usage of \NAME since deployment.}
	\label{fig:usage}
\end{figure}

\NAME has been used to develop, test and run a broad range of analytics on an array
of datasets. \NAME currently manages datasets that collectively 
total nearly 10TB of data. It has been used for varied development cycles ranging from researchers 
running off-the-shelf tools on sample data, 
to teams of programmers developing and testing
new methods on large, proprietary datasets.
Throughout its development, \NAME was designed to speed up development
and analysis cycles in a secure and flexible manner, while reducing costs
induced by disparate and often idle compute resources.
To-date, \NAME has been used to process over 5TB of data with over 75,330 CPU-hours.
As our implementation of \NAME has solidified and become more robust, usage has grown (see Figure
~\ref{fig:usage}).
The observed usage patterns affirm the choice of elastic cloud computing 
infrastructure as both data access and compute usage are 
particularly sporadic with peaks of over 7,000 compute hours in a single
day and other days with none.

\subsection{Applications}
\NAME currently hosts a number of proprietary and sensitive datasets that
have been used for a variety of workloads.
\NAME has been primarily used to develop and run time-intensive text analyses,
large scale matrix factorization, and optical character recognition (OCR). 
Other use cases have also been deployed with \NAMENS, but these three 
exemplify cases for which \NAME was designed: 
they are exploratory, large scale, and require private data 
that is shared among a strict set of users.

\subsubsection{Text Analysis}
One of the first use cases that was developed to run on \NAME
is a tool-chain for text analysis.
The analysis, designed to run on large collections of text, consists 
of four phases, each with separate inputs. 
The jobs consists of a series of pre- and post-processing scripts
and wrapping natural language models. For these jobs, data 
is normalized and divided into logical bins, submitted to semantic analysis
and post-processed to provide organized output.
The semantic model -- the central analysis --
includes doc2vec, word2vec~\cite{Mikolov} and various 
probabilistic topic models~\cite{Rosen-Zvi,Zhang}. These
tend to be memory and compute intensive.

\begin{figure}
  \centering
  \includegraphics[scale=0.43] {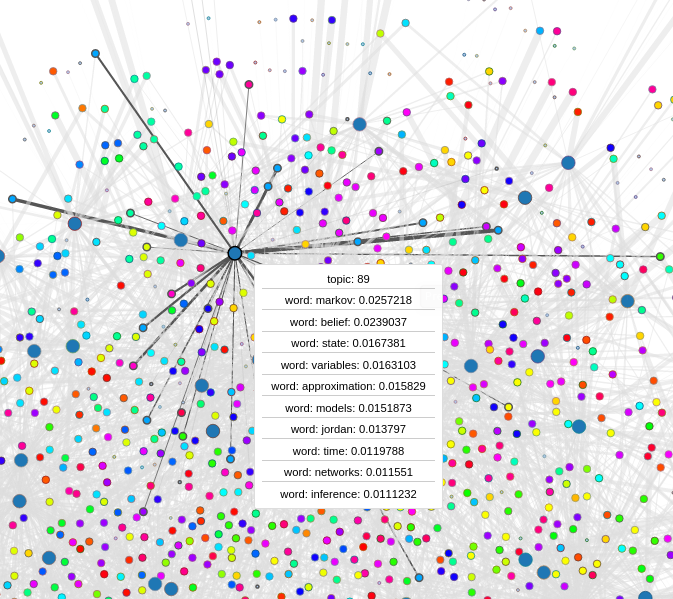}
  \caption{Interactive analytics based on the analysis of researchers' publications.
	Shown is the author-to-topic network highlighting connections via topic \#89, 
	about Markov models.}
    \label{fig:at_shots}
\end{figure}

This workflow has unique challenges that can discourage 
exploration and slow development. It tends to involves free 
parameters at various phases, the scaling profile of many
semantic models is unpredictable and the veracity of results 
is often measure qualitatively, making exploratory runs crucial.
\NAME allowed our users to efficiently
explore the parameter space to optimize a models' 
specification. One particular example is the Author-Topic 
(AT) Model~\cite{Rosen-Zvi}, a probabilistic graph model that 
fits distributions of words, documents and authors 
to topics observed in texts. With \NAMENS, a 
multi-dimensional grid search was performed to assess the 
quality of various specifications attenuating the number of topics, 
and various fitting parameters that affect the interpretability of topics.
Each run took approximately three days and 16GB of RAM.
The final outputs of these runs were used to develop an interactive 
platform for researchers (Fig. \ref{fig:at_shots}) to generate 
explore commonalities and to propose 
future collaborators based on their existing work~\cite{Gerow}. 

\subsubsection{Matrix Factorization}
Another use case where \NAME has been used was in
developing a new method of multiple imputation (MI).
When faced with lossy data, researchers often use 
MI to fill in missing values. Traditional MI establishes a 
missingness pattern on a given response 
which is then used
in a regression with non-missing 
responses as parameters. The ``multiple'' aspect of MI 
is that after being imputed, new values can be used to increase 
the accuracy of imputing still-missing values. MI is
cross-validated to assess stability and provide
error-bounds. As a result, MI is computationally intensive 
and imposes strong statistical assumptions. A more fundamental  
weakness of parametric MI, however, is that it disregards latent 
structure in the response matrix. Low rank and low 
norm matrix factorization offer alternatives to parametric 
MI that can exploit patterns throughout the response structure \cite{Udell}. 

In developing and testing low rank and low norm alternatives 
to MI, \NAME was used to run a range of tests 
simultaneously. These models were implemented in \textit{Julia},
a relatively new scientific programming language.
Because there is a stochastic component to low rank and low norm models, 
cross-validation on a held-out set of data was used to evaluate 
results over many random folds. Once the models were 
developed, \NAME was able to execute a large batch of 
validation jobs to provide pooled results. In these sets, a single
run on a 2,500 by 122 matrix used 32 cores on a single instance and 100 
GB of RAM for 10 hours. This made \NAMENS's ability to run multiple
jobs in parallel a crucial feature over the course of development.

\subsubsection{Optical Character Recognition}
A third application for which \NAME has been used in production 
is OCR. OCR is the process of extracting
text, figures, tables, and other features from rasterized
images of documents. OCR is effectively a kind of object recognition 
that relies on trained models of character classification -- models
that are computationally intensive. \NAME was used to run OCR software 
on over 10 thousand grant proposals and scholarly texts. Processing these
documents required 20 hours using 10, 32-core instances and 75GB 
of RAM. With \NAME, what would have taken over a month on personal hardware,
was finished in a single day.

\section{Related Work}\label{sec:relatedwork}



Computational social science communities are investigating
a broad range of approaches for hosting proprietary datasets
and conducting scalable analytics. For example, 
researchers have used hybrid cloud models~\cite{abramson14hybrid}, 
extended common tools to analyze data at scale~\cite{saleem14bigexcel}, 
and developed environments for securely analyzing data in controlled VMs~\cite{zheng14capsules}.
\NAME is unique in its support for a wide range of data, a general
architecture that accommodates many use cases, 
and by its automated, scalable storage and analytics environments. 

Many scientific communities now have a broad range of
data repositories available for storing and accessing
different types of data (e.g., biomedicine~\cite{dbgap}, climate~\cite{ncdc}, and astronomy~\cite{simbad}).
Systems are typically developed around a static data
repository that requires significant administrative overhead
to populate, curate and manage. Each has independent identity management
sub-systems that have been developed to control access to data.
But more importantly, most existing systems are
static, isolated data environments, that provide minimal
management capabilities separate from compute resources.
Science gateways~\cite{wikinsdiehr07gateways} aim to bridge
this gap by abstracting the complexity of using large scale computing
infrastructure.
These systems typically provide access to shared datasets (e.g., in a repository) and resources through 
high level interfaces (workflows, portals, etc.). 
Examples of commonly used gateways include CyberGIS~\cite{liu13cybergis} for
geoscience and iPlant~\cite{stanzione11iplant} for ecology.
Most science gateways are built on more traditional 
High Performance Computing (HPC) infrastructure. However,
recent work has focused on cloud-based solutions~\cite{wu11cloud, madduri2014globus}.
\NAME acts as a fabric on which next generation data repositories and cloud-hosted
gateways could be developed in a domain-agnostic setting. 


\NAME can deploy customized, cloud-based clusters similar
to a number of other systems. For example, 
CloudMan~\cite{cloudman} and StarCluster~\cite{starcluster}
allow uses to deploy clusters
for hosting and executing workflows.
These systems, and others, are designed to aid the creation
of clusters for semi-permanent usage.
Other systems, such as Globus Galaxies~\cite{madduri2014globus}
and Makeflow~\cite{albrecht12makeflow}, enable on-demand and elastic cluster  
provisioning in response to workload.
\NAME is unique, however, in
its use of commodity AWS services and its broad focus 
on providing a framework for secure data storage and analysis. 



\section{Summary}
\label{sec:summary}

\NAME provides a secure and scalable data enclave and analytics
environment for computational and data-drive social sciences. 
It addresses a gaping hole in the current infrastructure available
to researchers, providing a model via which, even resource limited 
researchers can gain access to scalable data storage, elastic
computing capacity, and cutting edge analysis algorithms without
deploying and operating their own infrastructure. 
Moreover, it provides these capabilities while also 
optimizing performance and cost using automated
data management and compute provisioning techniques. 

In the six months since deployment \NAME has quickly grown
to host a dozen private datasets (e.g., IEEE, Web of Science, and ACM)
as well as several public datasets (e.g., US patents and Wikipedia). 
It has been used by dozens of researchers, students, and teachers
to perform a wide variety of text analysis, machine learning, image
recognition, and network analysis algorithms. 

Our future work focuses on building an ecosystem around \NAME
by developing a suite of data analytics frameworks from which
users can more easily conduct analyses. We will continue to
engage social scientists to add datasets to the system while looking to extend its capabilities to other disciplines. 
We are particularly interested in further developing algorithms
for improving data lifecycle and compute management to better
meet the needs of users with respect to performance, time, and cost.

\section*{Acknowledgments}
The authors thank Nandana Sengupta, Nathan Bartley, and Cha Chen for developing applications on \NAME. 
This research was supported by grants from the John Templeton Foundation to the Metaknowledge Research Network, 
IBM for Computational Creativity, and a gift from Facebook.




\bibliographystyle{IEEEtran}
\bibliography{references}
%

\end{document}